\documentclass[debug]{epl}
\usepackage{epsfig}
\title{Raman Scattering in Charge Ordered Pr$_{0.63}$Ca$_{0.37}$MnO$_{3}$ :
Anomalous Temperature Dependence of Linewidth}
\shortauthor{R. Gupta {\it et al.}}
\shorttitle{ Anomalous Raman Linewidths in CO Manganites}
\author{Rajeev Gupta\inst{1}
\thanks{Present Address : 104 Davey Laboratory, Pennsylvania State
University, University Park, PA 16802, USA}, 
G. Venketeswara Pai\inst{2}
\thanks{Present Address : The Abdus Salam International Centre for
Theoretical Physics, Strada Costiera 11, 34100 Trieste, Italy}, 
A. K. Sood\inst{1}\thanks{ Email : [rajeev, venkat, asood, tvrama]
@physics.iisc.ernet.in ;  cnrrao@jncasr.ac.in },\\ 
T. V. Ramakrishnan\inst{2} \and C. N. R. Rao\inst{3}}
\institute{
\inst{1} Department of Physics, Indian Institute of Science,
Bangalore - 560012, India.\\
\inst{2} Centre for Condensed Matter Theory, Department of Physics,
Indian Institute of Science, Bangalore - 560012, India.\\
\inst{3} CSIR Centre for Excellence in Chemistry, Jawaharlal Nehru Centre
for Advanced Scientific Research, Jakkur P.O, Bangalore - 560064, India.
}
\begin{document}
\maketitle
\begin{abstract}
We report on the evolution of the Raman-active A$_g$ phonon modes in
the charge-ordered manganite Pr$_{0.63}$Ca$_{0.37}$MnO$_{3}$
as a function of temperature from 300K to 25K.
Our studies reveal that the linewidths of the A$_g$(2) and A$_g$(4)
phonons increase as temperature decreases.
This anomalous temperature dependence of phonon lineshapes,
seen for the first time in charge ordered manganites, can be
quantitatively understood in terms of a strong spin-phonon coupling
involving $t_{2g}$ spins and A$_g$ phonons.
\end{abstract}
The recent upsurge of interest in doped manganites exhibiting colossal
magnetoresistance
has fuelled a lot of interesting questions 
regarding the role of  spin, orbital and
lattice degrees of freedom and the interplay amongst them \cite{tokurareview}.
These systems of the form
$A_{1-x}A'_{x}$MnO$_{3}$, where $A$ is a trivalent rare earth ion 
({\it e.g.}, La, Pr, Nd) and $A'$ is a divalent ion
({\it e.g.}, Ca, Sr, Pb) show a rich phase
diagram depending on the tolerance factor and the amount of doping $x$ which in
turn controls the ratio of Mn$^{3+}$ to Mn$^{4+}$.
The essential ingredients for understanding their electronic properties
are the double-exchange(DE) mechanism \cite{zenerphil}
and the polaron formation due to the Jahn-Teller(JT) effect \cite{millis}.
In the case of systems where the weighted average 
$A$-site cation radius is small,
the system also shows charge ordering (CO),
{\it i.e.,} real space ordering of
Mn$^{3+}$ and Mn$^{4+}$ ions at low temperatures\cite{cnr}. The CO state
becomes stable when the repulsive Coulomb interaction between the carriers
or the cooperative JT effect
dominates over the kinetic energy of carriers.
In such situations there is
a strong competition between the DE interaction which favours
ferromagetism (FM), and CO which favours antiferromagnetism (AF) 
via the coupling between the background $t_{2g}$ ($S=3/2$) spins.

The system under study, Pr$_{1-x}$Ca$_{x}$MnO$_{3}$, shows a rich phase diagram
as a function of doping and temperature and has been  studied using a variety of
probes like resistivity \cite{tom1}, magnetization and
neutron diffraction \cite{lee1}. In the
temperature range T $>$ T$_{CO}$ ($\sim$ 240 K) the system is a paramagnetic
insulator. In dc magnetic susceptibility, a large peak is observed at T$_{CO}$
followed by a relatively small peak at T$_{N}$. The peak at
T$_{CO}$ is attributed to ferromagnetic correlations \cite{cox1}. The system
further undergoes a transition at T$_N$ ${\sim 175 K}$ to the
charge-exchange (CE) type
antiferromagnetic insulator (AFI). The CE type
structure for $x <$ 0.5 is different from what is present in systems with $x =
0.5$. In this so called "pseudo-CE" structure the zig-zag FM chains in the
$ab$ plane are FM aligned along the $c$-axis \cite{cox1}. This is unlike the CE
structure where the layers in the $ab$ plane are aligned antiferromagnetically
along $c$. Further lowering of temperature below 50 K leads to another
transition which can be understood either as a canted antiferromagnetic state
\cite{lee1} or as a mixture of ferromagnetic domains (or clusters) in an
antiferromagnetic background.

In this letter
we report our Raman studies on
well characterized, high-quality single crystals of
Pr$_{0.63}$Ca$_{0.37}$MnO$_{3}$ \cite{aguha} as a function of temperature 
from 300K to 25K. There are many Raman studies on manganites in recent times
covering both the vibrational\cite{ilie,irwin,dope1,yoon,gran2}
and electronic Raman scattering \cite{raj1}
across the metal-insulator transition.
In a recent work on polycrystalline Pr$_{0.65}$Ca$_{0.35}$MnO$_{3}$
 \cite{dediu},
temperature dependence of the modes at 475 and 610cm$^{-1}$ was
studied and the splitting of these modes below T$_{CO}$ was
attributed to the folding of the Brillouin zone in the CO state.
In another Raman study \cite{ganin}, the effect of $A$-site substitution on
the A$_g(2)$ mode was discussed. However, the linewidths have not been
discussed, and as we show in this letter, they reveal a very unusual
behaviour.
The aim of the present study of Pr$_{0.63}$Ca$_{0.37}$MnO$_{3}$ 
is to explore the coupling
between phonons, electrons and spins by examining the temperature dependence of
Raman linewidths across the CO and magnetic transitions.
Our studies do reveal  a strong
anomalous temperature dependence of the linewidths of 
the two Raman active A$_g$ modes,
which we have quantitatively analysed in terms of strong spin-lattice coupling
in these systems.

The single crystal used for the experiment was prepared by the float zone
technique \cite{tom1}. A (100) crystalline face of dimensions 5mm diameter
and 2mm thickness was polished using diamond paste and mounted on the cold
finger of the closed-cycle helium refrigerator (RMC model 22C CRYODYNE) using
thermally cycled GE (M/s. General Electric, USA) varnish. Raman spectra
were recorded in the spectral range 150 cm$^{-1}$ to 750
cm$^{-1}$ in  back-scattering geometry using a
DILOR XY spectrometer equipped with a liquid-nitrogen-cooled charge-coupled
device detector.
An argon ion laser line at 514.5nm was used as the
excitation wavelength at laser power of $\sim$ 25mW focussed  to a spot of
40$\mu$m on the sample. The temperatures
quoted are those measured on the cold finger using a Pt-sensor
coupled to a home-made temperature controller. The temperatures are within an
accuracy of $\pm$ 1K and do not take into account laser heating.

For an ideal perovskite $AB$O$_3$
with cubic structure (space group O$_{h}^{1}$, $Pm3m$) the irreducible
representations at the zone center are given by 4F$_{1u}$ + F$_{2u}$. The only
allowed optical modes are the infrared modes and since all lattice sites have
inversion symmetry, first-order Raman scattering is forbidden.The small size
of the Pr cation in comparison to the oxygen causes a large tilting of the
MnO$_{6}$ octahedra leading to a GdFeO$_{3}$ structure.
The space group of Pr$_{0.63}$Ca$_{0.37}$MnO$_{3}$ at room
temperature is $Pbnm$ (D$^{16}_{2h}$); there are four formula units per
unit cell and there are 24 Raman active modes \cite{ilie}.
Iliev {\it et al.} \cite{ilie} have carried out
systematic Raman measurements at room temperature and lattice dynamical
calculations on undoped, stoichiometric, orthorhombic YMnO$_3$ and LaMnO$_3$
with $Pbnm$ structure and identified the symmetry of the different modes.
The Raman modes are weak even in stoichiometric LaMnO$_{3}$; the
observed modes at 140, 198, 257, 284 and 493cm$^{-1}$ were assigned
A$_g$ symmetry, modes at 109, 170, 308, 481 and 611cm$^{-1}$ as B$_{2g}$ and
modes at 184 and 320cm$^{-1}$ as B$_{1g}$ or B$_{3g}$.
Recent experiments on LaMnO$_3$ have revealed the presence of  high-energy modes
($\sim$ 1000cm$^{-1}$) which have been attributed to the orbital excitations
\cite{saitonature}. However, doped manganites being pseudo-cubic,
Raman lines are extremely weak and broad \cite{ilie,irwin,dope1,yoon,gran2}.
The number of modes observed in doped manganites are very few; {\it e.g.}, in
La$_{0.7}$Ca$_{0.3}$MnO$_{3}$ there are five modes observed at low
temperature at 70, 133, 235, 435, and 670cm$^{-1}$ \cite{dope1}.

\begin{figure}
\twofigures[height=10cm]{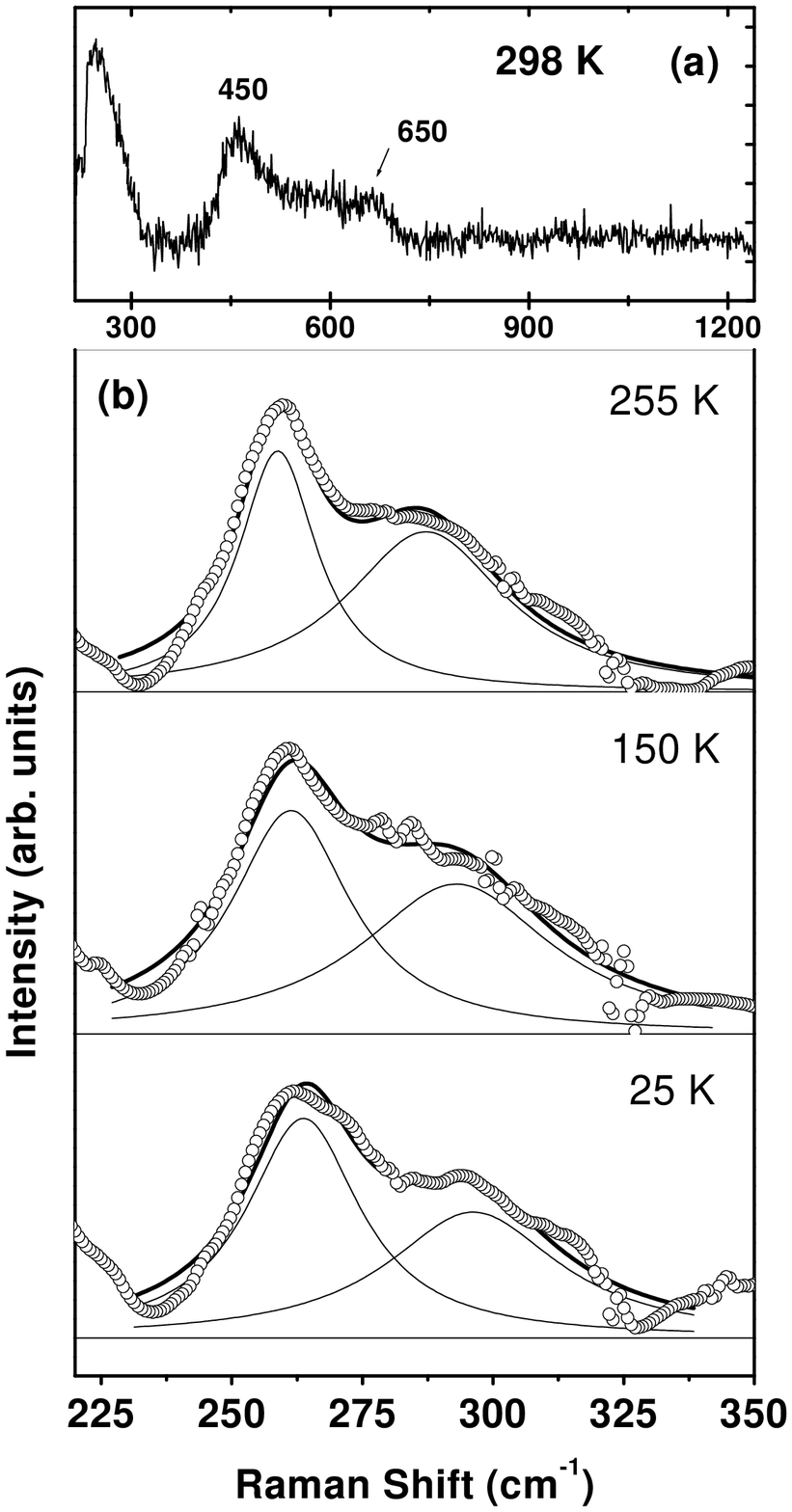}{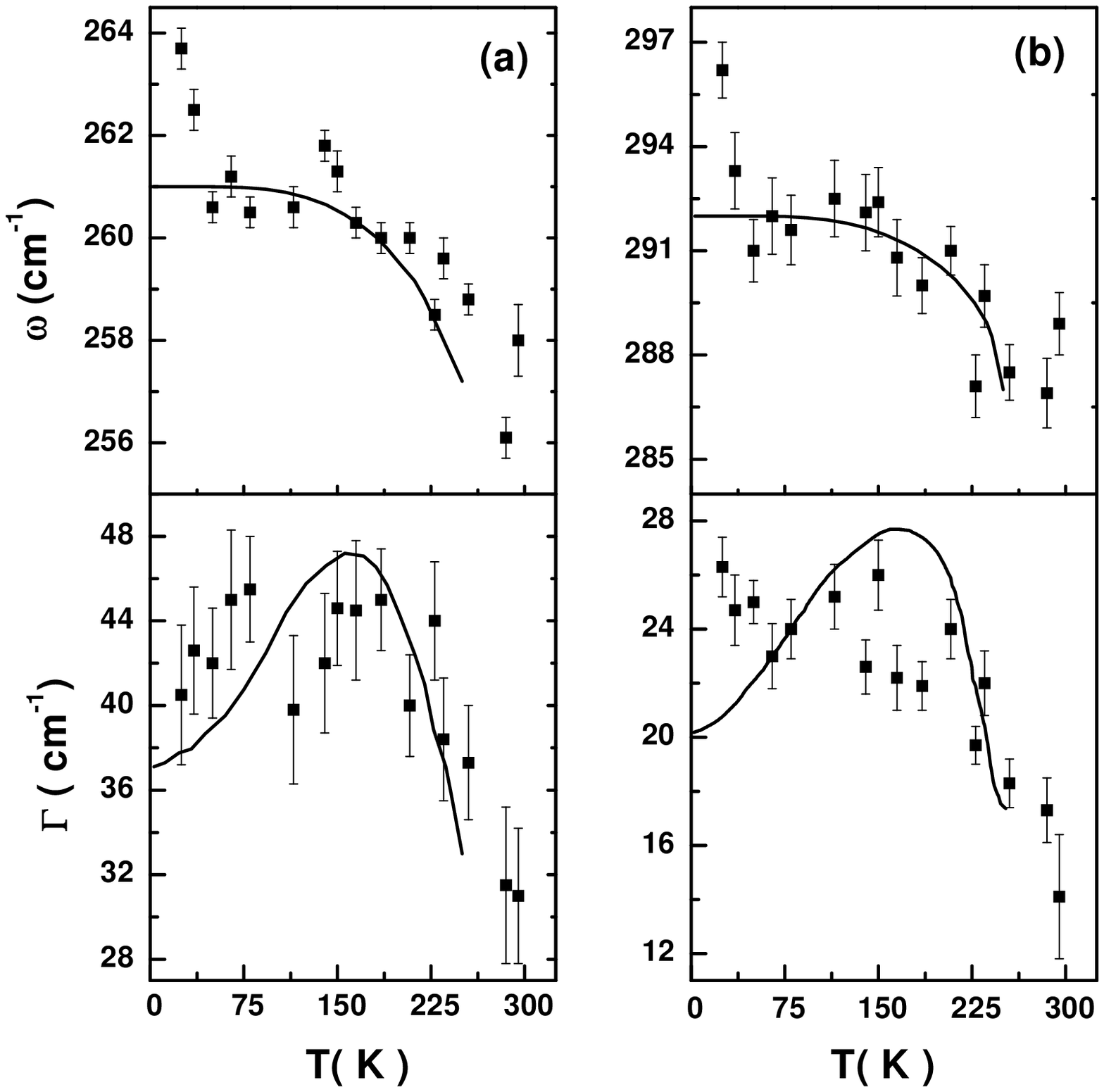}
\caption{
Raman spectra as a function of temperature.
(a) Micro-Raman spectrum taken at 298 K showing the modes at $\sim$
260, 450, and 650 cm$^{-1}$. (b) Raman spectrum of A$_g$(2) and
A$_g$(4) modes at three different temperatures. The thick solid smooth
curves show the sum of the two fitted Lorentzian lineshapes (thin
solid smooth curve). Raw data has been denoised using a wavelet filtering routine (Daubechies wavelet filter of order 4 at level 4).}
\caption{ Variation of the Raman frequency ($\omega$) and the
full width at half-maximum ($\Gamma$) as a function of temperature.
Filled squares are the experimental values and thick solid curves
are the theoretical results in our model calculation.
Panel (a) is for the A$_g$(2) mode and panel (b) is for the A$_g$(4)
mode.}
\end{figure}

In the micro-Raman setup using 6328 \AA  radiation from He-Ne laser,
we observe four bands
centered around $\sim$ 260, 290, 450 and 650cm$^{-1}$.
However,  in macro-geometry with the crystal in the cryostat,
only two modes centered around 258 and 288cm$^{-1}$ at 300K
were observed with sufficient signal to noise ratio (Fig. 1).
The other bands near 450 and 650cm$^{-1}$ are weaker and therefore
could not be followed as a function of temperature.

We will now discuss the origin of the observed modes. Since the crystal
structure for Pr$_{1-x}$Ca$_{x}$MnO$_{3}$ is same as that of LaMnO$_3$, it is
plausible to assign to  these modes the symmetry used by Iliev {\it et al.}
\cite{ilie} for LaMnO$_3$. The low-frequency 
mode at 258cm$^{-1}$ has been attributed to the A$_g$(2) mode which is
the in-phase rotation of the oxygen cage about the $y$-axis on adjacent MnO$_6$
octahedra. In this mode only the in-plane oxygen atoms, 
namely O1, are involved. The
mode at 288cm$^{-1}$ has been assigned to the A$_g$(4) mode
which is out-of-phase rotation of the oxygen cage about the $x$-axis. 
Here along with the in-plane oxygens (O1), the apical oxygens (O2)
are also involved. The Mn ion being at the centre of
inversion symmetry remains stationary in both the modes.
Figs. 2a and 2b show the variation of the peak position and linewidths (full
width at half-maximum) for the two modes as a function of temperature,
obtained by nonlinear least-square fitting of the data with a sum of two
Lorentzians along with a baseline. The mode frequencies increase as the
temperature is lowered and a sharp change is noticeable at
$\sim$ 50 K as the system undergoes a transition
to a canted AFI phase \cite{tom1}. Most interestingly, the
linewidths for both the modes show anomalous behaviour in that the widths
decrease as the temperature is increased. Another notable observation is the
large value of linewidth even at low temperatures.

Next we look at the various possibilities that may lead to the observed 
temperature dependence of mode frequencies.
Mode softening ($\sim$ 3\%) as the temperature
is raised can arise due to the
quasi-harmonic contribution arising from the lattice
expansion which can, in turn, change the force constants. This change in
frequency is related to the Gr\"uneisen parameter and the volume change.
Since the fractional volume change across the studied
temperature regime is not available for Pr$_{0.63}$Ca$_{0.37}$MnO$_{3}$
we can be guided by the data available in literature for
Pr$_{0.7}$Ca$_{0.3}$MnO$_{3}$.
The volume change is
$\sim$ 0.4\% between temperatures of 300 and 10 K. This change is too small to
account for the observed frequency changes of both the A$_g$ modes. 
Another contribution can
arise due to the intrinsic anharmonic (cubic) interactions.
This contribution is also insufficient to explain the large
softening, since typically for perovskites, {\it e.g.,} KTaO$_{3}$,  the
change of peak position is about  2-3 cm$^{-1}$ in the
temperature range 10-500 K \cite{perry}.

Now we consider the linewidths.
Anharmonic interactions  will lead to
an increase of the linewidth with increasing temperature,
in contrast to our experimental results.
Lee and Min \cite{lee2} have proposed a model to explain the sound velocity
softening across the charge ordering transition in
La$_{0.5}$Ca$_{0.5}$MnO$_{3}$  \cite{sound}.
Recently Dattagupta and Sood \cite{datta1} have extended this model
incorporating both the DE interaction and electron-optical phonon coupling.
Their models also fail to explain the behaviour of linewidth at low
temperatures in the magnetically ordered phase.
We suggest that magnon-phonon interaction
leads to the shifts in phonon frequencies  as well as
the observed temperature dependence of
lifetimes. The
frequency shift can be understood as arising due to the modulation of the
exchange integral between
Mn ions. This has been
considered earlier for
LaMnO$_{3+\delta}$ \cite{gran2} 
to explain the softening of the Mn-O-Mn stretching mode.
However, this model also does not take into account the dynamics
and hence cannot explain the temperature dependence of
the linewidth($\Gamma$). 
A sharp increase in the peak positions at low temperatures near the
canted AF transition (see Fig. (2)) suggests a strong coupling
between the Raman modes and spin excitations, which we show to be
also responsible for the anomalous temperature dependence of the linewidths.

We briefly outline here the origin of the large spin-phonon coupling
and the way it leads to anomalous phonon damping.
Details will be published elsewhere \cite{venkat2}.
When the neighbouring MnO$_6$
octahedra are tilted, the Mn-O-Mn bond angle
$\theta_{ij} (= [\pi-(\theta_i-\theta_j)])$ between nearest-neighbour Mn-O
bonds involving Mn ions $i$ and $j$ changes. Here $\theta_i$ and
$\theta_j$ are the small angles of rotation of the neighbouring octahedra
centred at $i$ and $j$, respectively. Since the oxygen mediated overlap
$t_{ij}$ between Mn-$d$ orbitals goes as $\cos \theta_{ij}$,
$J_{ij}$ goes as  $J_{ij}^0 \cos \theta_{ij}/2
\sim J_{ij}^0 (1- {(\theta_i-\theta_j)^2 \over 2})$. This has a term
of the form $\theta_i \theta_j$ which, among others, corresponds to
a simultaneous excitation of the
rotation mode involving the Mn-O bond, with a displacement
$q_{1i}(\sim a\theta_i)$ and another mode $q_{2j}(\sim a\theta_j)$
where $a$ is the Mn-O bond length.
The coupling between the two phonons will be strongest 
if they belong to the same symmetry
representation. The first of these is the Raman-active phonon 
under consideration,
with a frequency $\omega_1 =$ 258cm$^{-1}$ ($A_{g}(2)$) or 
$\omega_1 =$ 288cm$^{-1}$ ($A_{g}(4)$)
and the other is an optical
phonon mode with frequency $\omega_2$.

For simplicity in calculation, we assume that 
the magnetic order is of the C-type in which
ferromagnetic chains are linear in contrast to the zig-zag chains
in the CE structure. Both these structures are geometrically similar,
have the same number of nearest-neighbour FM and AF bonds per spin,
and have the same magnetic energies per spin \cite{venkateldoped}. 
Though there can
be quantitative differences in their magnetic excitation spectrum,
qualitatively both are very similar in that they have mixed
FM and AF character along arbitrary directions
of the Brillouin zone with the same number of nearest-neighbour
FM/AF couplings and the same overall geometry, namely FM along a line and
AF in plane. The latter decides the dispersion relation and the phase
space for the mixed FM and AF spin excitations.
Since we are interested in understanding how
magnetic fluctuations affect the phonon spectrum, it is therefore a very good
approximation to work with the C-type structure rather than the CE or
pseudo-CE type structures.
The effective spin-Hamiltonian for this anisotropic spin system is given by
\begin{eqnarray}
H_{spin}&=&\sum_{\left<ij\right>;i,j \in xz}J_{AF}^{xz}{\bf S}_i \cdot {\bf S}_j
- \sum_{\left<ij\right>;j \in y}J_F^y {\bf S}_i \cdot {\bf S}_j,
\end{eqnarray}
where $AF$ and $F$ refer to antiferromagnetic and ferromagnetic interactions.
Our Hamiltonian is most naturally
viewed as an effective Hamiltonian with $t_{2g}$ spin degrees of freedom
kept explicitly, and all other degrees of freedom (such as $e_g$ electrons,
JT coupling between $e_g$ electrons and lattice modes {\it orthogonal}
to those considered here, Hund's rule coupling) integrated out. (This is
possible and sensible because the $t_{2g}$ spins are the lowest energy
degrees of freedom near and below T$_N$, as discussed below.)
It has been shown
by several authors, {\it e.g.}, Kugel and Khomskii \cite{khomskii},
that in LaMnO$_3$
(which is an insulator experimentally showing JT distortions
and ferromagnetic planes coupled antiferromagnetically) the
Mn-O-Mn-mediated magnetic interaction is {\it ferro}magnetic in plane and
antiferromagnetic (usual superexchange) perpendicular to it.
The latter is indeed due to the same sort of virtual processes involving
the Mn-O-Mn bond as the AF superexchange. 
It has been shown that \cite{khomskii}
the JT distortions (and consequent $e_g$ orbital
excited states), and the Hund's rule coupling $J_H$ (which enslaves the
$e_g$ spins to the $t_{2g}$ spin in the ground state {\it and} produces an
excited state an energy $J_H$ away, with $e_g$, $t_{2g}$ spins
antiparallel) lead to a {\it ferromagnetic} Mn-O-Mn-mediated in-plane exchange
coupling. Thus detailed models involving DE and JT
coupling show as briefly mentioned above, that both AF and F coupling 
between $t_{2g}$ spins necessarily involve Mn-O-Mn overlap. Further, if there
are no other low-energy degrees of freedom ({\it e.g.}, no $e_g$ fermionic
low-energy metallic excitations, or orbital fluctuations), then the form of
Eq. (1) of the text with the discussed dependence on Mn-O-Mn angle, is
{\it inevitable}. A microscopic theory will relate the $J_{ij}$
to basic parameters of the
electron Hamiltonian, but $J_{ij}$ is always second order in the Mn-O-Mn
overlap. The only other possible low-energy degrees of freedom in the
{\it insulating phases} of CMR oxides are orbital fluctuations. Here
we contend that these are not relevant in the temperature range we are working
in because orbital order is already well developed.
For small fluctuations of the Mn-O bond angle, from the above discussion,
the spin-phonon interaction Hamiltonian ($H_{s-p}$) becomes
\begin{eqnarray}
H_{s-p} &=& -{g \over {2a^2}}\sum_{\left<ij\right>;i,j \in xz}
J_{oAF}^{xz}{\bf S}_i \cdot 
{\bf S}_j q_{1i}q_{2j}
+{g \over {2a^2}} \sum_{\left<ij\right>;j \in y}
J_{oF}^y {\bf S}_i \cdot {\bf S}_j
q_{1i}q_{2j} \nonumber\\
&+&\sum_{\left<ij\right>;i,j \in xz}
J_{oAF}^{xz}{\bf S}_i \cdot {\bf S}_j
- \sum_{\left<ij\right>;j \in y}
J_{oF}^y {\bf S}_i \cdot {\bf S}_j,
\end{eqnarray}
where $g$ is a coupling constant of order unity. The spin degrees of freedom
are described in terms of Schwinger bosons (SB)  \cite{hrk} which is an exact 
representation. It is
convenient both below and above the N\'{e}el temperature (T$_N$).
Below T$_N$, the SB Bose condense.
The spin-spin interaction in Eq. (2)
can be written in terms of boson operators $a_{k \sigma}^{\dag}$ in
mean-field theory as
\begin{eqnarray}
H_{SB}&=&H_0 + \sum_{k, \sigma} \omega_k[a_{k \sigma}^{\dag} a_{k \sigma} +1].
\end{eqnarray}
Here $H_0$ is a c-number (with no operators in it unlike the
second term which is the Hamiltonian of noninteracting SB)
to be determined self-consistently from the free energy, and
$\omega_k$ is the boson frequency,
\begin{eqnarray}
\omega_k &=& \sqrt{[(\lambda-J_{oF}^y z_y B_y \gamma_k^y)+\lambda]^2-
(J_{oAF}^{xz} z_{xz} B_{xz} \gamma_k^{xz})^2}.
\end{eqnarray}
Here, $B_{xz}$ and $B_y$ are the Weiss fields 
(the SB order parameters), and
$\lambda$ is the temperature dependent stiffness of the SB excitations, 
to be determined self-consistently from the free energy;
$z_{xz}(z_y)$ is the $xz$ plane($y$-axis) coordination number,
and $\gamma_p =1/z_p \sum_{\delta} e^{i {\bf k}_p.
{\bf \delta}}$ ($p=xz$ or $y$).
We choose $J_{oAF}^{xz} \sim$ 3.9meV and $J_{oF}^y \sim$ 7.8 meV so as to give
T$_N$ = 175 K.

The spin-phonon interaction in Eq. (2), written in terms of the SB
operators, leads to a coupling between the
A$_g$ mode being observed, and the second optical phonon mode of
wavenumber $\omega_2$, with the excitation of nearest-neighbour
spin fluctuations or boson pairs.
We calculate the decay rate for the A$_g$ phonons of frequency
$\hbar \omega_1$ in perturbation theory, to second order in $H_{s-p}$.
The phonon decays due to the second-order process which involves
the A$_g$ phonon decaying into the second phonon and two Schwinger bosons.
The imaginary (real) part of the self-energy $\Sigma$ determines
the decay rate (peak shift) of the Raman-active phonon.
Assuming the phonon modes to be dispersionless, the
imaginary part of $\Sigma$ is given by
\begin{eqnarray}
Im\;\: \Sigma &=& {g^2 \over {8 \pi a^2}} \sum_{q,k} f(q,k) 
\delta(\omega_1-\omega_2-\omega_k-\omega_q) [1+ g_{\omega(q+k)}+g_{\omega_{2}}]
[1+g_{\omega(k)}+g_{\omega(q)}],
\end{eqnarray}
where $f(q,k)$, the form factor is given by
\begin{eqnarray}
f(q,k) &=& {(J_{oF}^y B_y)}^2 p_1 A_1^2 + (J_{oF}^y B_y)(J_{oAF}^{xz}B_{xz})
A_1(p_2A_2 + p_3A_3)/2 \nonumber \\
&+& {(J_{oAF}^{xz}B_{xz})}^2(p_4 A_3^2 + p_5 A_2^2
+p_6 A_2 A_3)/4,
\end{eqnarray}
$p_i$ are products of Bogoliubov or coherence factors,
$g_{\omega(k)}^{-1} = e^{\beta \omega_k}-1$,
$A_1 = 2(\cos q_y + \cos k_y)$, $A_2 = 2(\cos q_x + \cos q_z)$, and
$A_3 = 2(\cos k_x + \cos k_z)$.

We find substantial linewidth at $T=0$, 
which initially increases with temperature
because of thermal availability of two SB states at energy
$\hbar (\omega_1 - \omega_2)$. At high temperatures, of order T$_N$ or higher,
the spin fluctuation mode softens, 
and the density of pair excitations at a fixed
energy difference $\hbar (\omega_1 - \omega_2)$ decreases as temperature
increases. The Weiss fields also decrease with increasing temperature.
Thus the A$_g$(2) phonon decay rate falls. This decrease has been
estimated in the SB formalism both below and above  T$_N$ (the
Bose condensation temperature in that formalism), or in a classical
theory for spin dynamics of a Heisenberg spin system specially useful
above T$_N$; the results above T$_N$ are roughly the same.
In the latter theory, one essentially obtains the nearest-neighbour
spin-fluctuation frequency spectrum via classical equations of motion
of coupled spin vectors. We show the Schwinger boson results in Fig. 2
for both linewidths and peak positions.
Guided by the lattice dynamic studies of Iliev {\it et al.} \cite{ilie}
on LaMnO$_3$, we have taken the A$_g$ phonon at 190cm$^{-1}$
as $\omega_2$.
The qualitative behaviour
of $\Gamma$ and peak postions will not change with another choice of $\omega_2$.
As can be seen, the agreement between the experiment 
and theory is rather good above 70K.
Our approach does not take into account the observed canted AFI state
below $T < 50K$, which may be  responsible for the discrepancy between the
theory and experiments at low temperatures. The canting of spins leads
to FM correlations in an otherwise perfectly ordered
AF state which, in turn, modifies the spin excitation
spectrum.

In conclusion, our Raman measurements on Pr$_{1-x}$Ca$_{x}$MnO$_{3}$ ($x$ =
0.37) reveal a strong coupling between the spin and lattice degrees of
freedom in CO  manganites. The large variation of the mode frequency
and anomalous behaviour of the linewidths as a function of temperature are
understood in terms of spin-phonon interaction. Thus, we believe, our
study reveals the striking consequences of  the strong interplay
between the spin and lattice degrees of freedom in CO manganites.

RG and GVP thank the Council of Scientific and Industrial Research (India) and 
AKS  thanks the Department of Science and Technology (India) for 
financial support.

\end{document}